\documentclass[twocolumn]{aastex701}

\usepackage{graphicx}
\usepackage{amsmath}
\usepackage{amssymb}
\usepackage{hyperref}
\usepackage{xspace}
\usepackage{booktabs}
\usepackage{longtable}
\usepackage{tikz}
\usetikzlibrary{shapes.geometric, arrows.meta, positioning, fit, backgrounds, calc}

\newcommand{\wtp}{WTP\xspace}
\newcommand{\unwise}{unWISE\xspace}
\newcommand{\neowise}{NEOWISE\xspace}

\newcommand{\sextractor}{SExtractor\xspace}
\newcommand{\psfex}{PSFEx\xspace}
\newcommand{\psql}{PostgreSQL\xspace}

\begin{document}

\title{Fifteen years of the variable infrared sky with the WISE Transients Project (WTP): Processing architecture and public data release}

\author{Kishalay De} 
\affiliation{Department of Astronomy and Columbia Astrophysics Laboratory, Columbia University, New York, NY, USA}
\affiliation{Center for Computational Astrophysics, Flatiron Institute, New York, NY, USA}
\email{kd3038@columbia.edu}

\author[]{Robert Simcoe}
\affiliation{MIT-Kavli Institute for Astrophysics and Space Research, Cambridge, MA, USA}
\email{}

\author[]{Aaron M. Meisner}
\affiliation{National Optical-Infrared Astronomy Research Laboratory, National Science Foundation, Tucson, USA}
\email{}

\author[]{Megan Masterson}
\affiliation{Center for Astrophysics | Harvard \& Smithsonian, Cambridge, MA, USA}
\email{}

\author[]{Lynne Hillenbrand}
\affiliation{Department of Astronomy, California Institute of Technology, Pasadena, CA, USA}
\email{}

\author[]{Christos Panagiotou}
\affiliation{MIT-Kavli Institute for Astrophysics and Space Research, Cambridge, MA, USA}
\email{}

\author[]{Viraj R. Karambelkar}
\altaffiliation{NASA Hubble Fellow}
\affiliation{Department of Astronomy and Columbia Astrophysics Laboratory, Columbia University, New York, NY, USA}
\email{}

\author[]{Geoffrey Mo}
\affiliation{The Observatories of the Carnegie Institution for Science, Pasadena, CA, USA}
\affiliation{Department of Astronomy, California Institute of Technology, Pasadena, CA, USA}
\email{}

\author[]{Anna-Christina Eilers}
\affiliation{MIT-Kavli Institute for Astrophysics and Space Research, Cambridge, MA, USA}
\email{}

\author[]{Danielle Frostig}
\affiliation{Center for Astrophysics | Harvard \& Smithsonian, Cambridge, MA, USA}
\email{}

\author[]{Jacob E. Jencson}
\affiliation{IPAC, California Institute of Technology, Pasadena, CA, USA}
\email{}

\author[]{Erin Kara}
\affiliation{MIT-Kavli Institute for Astrophysics and Space Research, Cambridge, MA, USA}
\email{}

\author{Ana Lam} 
\affiliation{Department of Astronomy and Columbia Astrophysics Laboratory, Columbia University, New York, NY, USA}
\email{}

\author[]{Ryan M. Lau}
\affiliation{IPAC, California Institute of Technology, Pasadena, CA, USA}
\email{}

\author[]{Gene Leung}
\affiliation{MIT-Kavli Institute for Astrophysics and Space Research, Cambridge, MA, USA}
\email{}

\author[]{Ashish Mahabal}
\affiliation{Department of Astronomy, California Institute of Technology, Pasadena, CA, USA}
\email{}

\author[]{Sam Rose}
\affiliation{Department of Astronomy, California Institute of Technology, Pasadena, CA, USA}
\email{}

\author[]{Kohki Uno}
\affiliation{Department of Astronomy and Columbia Astrophysics Laboratory, Columbia University, New York, NY, USA}
\email{}

\author{Justin Vega} 
\affiliation{Department of Astronomy and Columbia Astrophysics Laboratory, Columbia University, New York, NY, USA}
\email{}

\begin{abstract}
The Wide-field Infrared Survey Explorer (WISE) surveyed the entire sky at 3.4 and 4.6 $\mu$m ($W1$ and $W2$) every six months between 2010 and 2024, with a hiatus during 2011-2013. Despite its potential for IR variability studies, previous studies have primarily focused on variability of known sources discovered at other wavelengths. We present the WISE Transients Project (WTP), a systematic reprocessing of WISE observations using image subtraction techniques to transform WISE into a transient discovery machine. Using unWISE coadds, we construct reference images from the early mission (2010-2014) and subtract them from the complete set of time-resolved coadds. We detect variable and transient sources in the resulting difference images and generate alert packets analogous to modern synoptic surveys. Source-injection and recovery simulations demonstrate astrometric accuracy of $<0.5$\,\arcsec\ down to $\approx 19.5$\,AB\,mag, improving to $<0.05$\,\arcsec\ for sources brighter than $\approx 16$\,AB\,mag, where photometric accuracy reaches $\lesssim 0.05$\,mag. The 50\% detection completeness limits are $W1 \approx 19.6$ and $W2 \approx 18.8$\,AB\,mag, with reduced efficiency near regions of high source density. We publicly release the catalog containing $\approx 50$\,M alerts with i) cutouts of science, reference and difference images, ii) complete forced-photometry history over the mission ($\approx 4.6$\,B measurements), iii) a real-bogus classification score to identify point sources and iv) crossmatches to the CatWISE and 2MASS catalogs. WTP has already enabled a foundational census of the dynamic infrared sky, and this public catalog serves as a foundation for upcoming surveys by the Rubin Observatory and the Roman Space Telescope.
\end{abstract}


\section{Introduction}\label{sec:intro}

Over the past two decades, wide-field optical surveys have transformed time-domain astronomy. Facilities such as the Zwicky Transient Facility (ZTF; \citealt{Bellm2019b}), the Asteroid Terrestrial-impact Last Alert System (ATLAS; \citealt{Tonry2018}), and the All-sky Automated Survey for Supernovae (ASASSN; \citealt{Shappee2014}) routinely monitor large fractions of the sky, with the Rubin Observatory poised to extend this capability to substantially greater depths \citep{Ivezic2019}. Yet optical surveys provide an incomplete view of the dynamic universe. Transients in heavily obscured environments can escape detection, while phenomena embedded in circumstellar material may radiate predominantly in the infrared (IR). IR time-domain surveys therefore offer access to populations and phases of evolution that are difficult to identify at optical, ultraviolet, and X-ray wavelengths.

\begin{figure*}[!ht]
    \centering
    \includegraphics[width=0.99\linewidth]{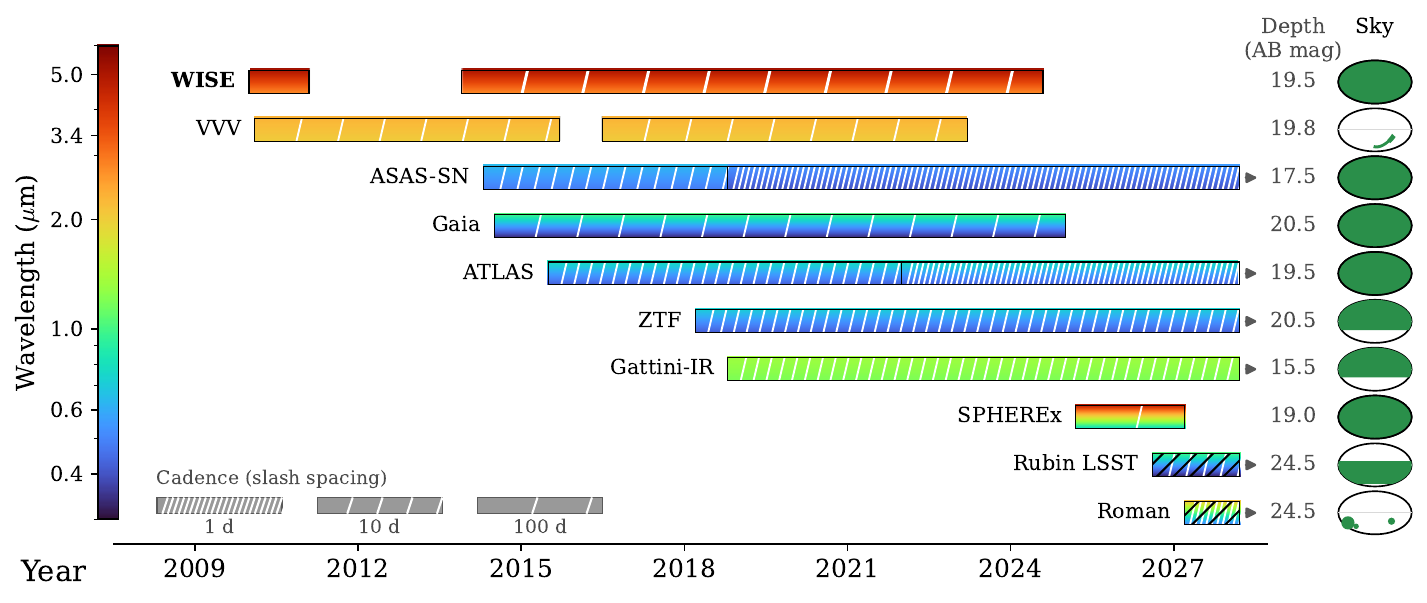}
    \caption{Schematic comparison of WISE/\neowise{} with optical and IR surveys of the past and coming decade that repeatedly scan the sky (ordered by survey start date) -- VVV \citep{Minniti2010}, ASASSN \citep{Shappee2014}, Gaia \citep{Gaia2016}, ATLAS \citep{Tonry2018}, ZTF \citep{Bellm2019}, Gattini-IR \citep{De2020}, SPHEREx \citep{Bock2026}, LSST \citep{Ivezic2019} and the Roman Space Telescope. Horizontal bars indicate survey operation timelines, color gradients represent wavelength coverage, and the spacing of diagonal marks illustrates representative revisit cadences, as shown in the lower-left legend. Arrows indicate continuing or planned survey operations beyond the plotted interval. The right-hand columns give representative depths in AB magnitudes and schematic sky footprints.}
    \label{fig:landscape}
\end{figure*}

Large area surveys at IR wavelengths began with repeated IR monitoring of substantial regions of the Milky Way. These include targeted surveys of star forming regions with the 2MASS \citep{Carpenter2001, Carpenter2002} and Spitzer telescopes \citep{YSOVar}. Subsequently, the VISTA Variables in the Via Lactea survey (VVV; \citealt{Minniti2010}), covered $\approx 520$\,sq.\,deg. in the $K_s$ band, accumulating approximately 100 epochs over a decade. On the other hand, early extragalactic IR transient searches largely used pointed observations to target nearby galaxies with \textit{Spitzer} \citep{Kasliwal2017, Boyer2015} or used ground-based observations to uncover obscured supernovae in luminous IR galaxies \citep{Kool2018, Mattila2007}.

The Palomar Gattini-IR (PGIR; \citealt{De2020}) survey established routine monitoring of the entire accessible sky in the $J$ band, covering $\approx 15000$\,sq.\,deg. at an approximately two-night cadence to a depth of $\approx 15.5$\,AB\,mag. Thanks to innovations in detector technology and real-time data processing, the Wide-field Infrared Transient Explorer (WINTER; \citealt{Lourie2020, Frostig2025}) leverages a 1\,sq.\,deg. camera to survey large selected regions, including the Galactic plane \citep{Frostig2026} and the Andromeda galaxy \citep{Karambelkar2025}. More recently, the Prime-focus Infrared Microlensing Experiment (PRIME; \citealt{Sumi2025}) has begun monitoring the Galactic bulge at high cadence, with several visits per night, as a precursor to the Roman Galactic Bulge Time Domain Survey.

Although originally designed for creating a static map of the infrared sky, the \textit{Wide-field Infrared Survey Explorer} (WISE; \citealt{Wright2010}) provides a uniquely powerful resource for IR time-domain astronomy. WISE and its reactivated \neowise{} mission (aimed at characterizing solar system objects; \citealt{Mainzer2014}) observed the sky at 3.4 and 4.6\,$\mu$m ($W1$ and $W2$) between 2010 and 2024, with a hiatus during 2011--2013. Most sky locations were revisited approximately every six months, yielding repeated observations over a nearly 15-year baseline. These data provide uniform all-sky coverage with sensitivity at wavelengths where the thermal background makes comparable ground-based surveys impractical. Figure~\ref{fig:landscape} places this archive in the broader survey landscape, illustrating its unmatched combination of temporal baseline, wavelength coverage, and sky coverage relative to other time-domain facilities.

Owing to its slow cadence, which is not optimal for discovering extragalactic transients, much of the time-domain work using WISE/\neowise{} has focused on retrospective studies of sources identified at other wavelengths. Examples include searches for dust echoes from tidal disruption events \citep{vanVelzen2016}, variability in young stellar objects \citep{Hillenbrand2018} and Active Galactic Nuclei \citep{Jiang2021}, and dust emission from supernovae \citep{Thevenot2020}. These studies demonstrate the scientific reach of the archive, but searches conditioned on discoveries at other wavelengths can miss sources whose detectable signatures are confined to the IR. An untargeted census of such phenomena requires a systematic search for variability directly in the mid-IR images.

On the technical front, nearly all of these previous variability searches in NEOWISE data have leveraged directly accessible NEOWISE source catalogs made available as part of the annual data releases (e.g., \citealt{Paz2026, Kim2026, Petrosky2021}); however, source confusion limits their utility for systematic time-domain searches without image subtraction \citep{Lang2014}. The \texttt{untimely} catalog \citep{Meisner2023} mitigates some of the source confusion limitations introduced by the official WISE data products, but without image subtraction necessary for the most crowded fields (e.g. in the Galactic plane and on top of nearby galaxies) given the coarse size of the WISE point spread function (PSF; $\approx 6$\,\arcsec). 

To this end, we initiated the WISE Transients Project (\wtp{}) as a dedicated effort to discover transients and variables directly in WISE data by processing its archival images with the computational infrastructure of a modern synoptic time domain survey. Much of the foundation for this effort was laid by the \unwise{} project \citep{Lang2014,Meisner2018}, which combined individual WISE/\neowise{} exposures into coadds on a fixed sky grid in temporally resolved bins. WTP applies image subtraction techniques established by modern synoptic surveys \citep{Masci2019, De2020} to the \unwise{} coadds, performs source detection on difference images to identify variable sources and creates an archive of `alerts' stored in relational databases to enable rapid searches of the archive. 

The WTP approach has already enabled a diverse set of discoveries that would otherwise have been inaccessible -- including a promising planetary engulfment event \citep{De2023}, heavily obscured Galactic transients such as novae \citep{Zuckerman2023} and young stellar outbursts \citep{Tran2024, Frostig2026, Karambelkar2026}, dust echoes of Galactic supernovae \citep{Vega2026}, the nearest known tidal disruption event \citep{Panagiotou2023}, a broader population of dust obscured tidal disruption events \citep{Masterson2024, Nair2026} and searches of their multi-messenger counterparts \citep{Wang2026}, a failed-supernova candidate \citep{De2026} and the onset of a common envelope event \citep{Karambelkar2025, Karambelkar2026c} in the Andromeda galaxy, and populations of nearby supernovae with luminous late-time IR emission associated with circumstellar interaction \citep{Myers2024, Uno2026}. Together, these results demonstrate the value of mid-IR variability datasets and motivate making this capability available to the wider community.

In this paper, we present the WTP processing architecture, characterize its performance, and publicly release the resulting alert catalog. Section~\ref{sec:pipeline} describes the input data to the pipeline, processing architecture and image subtraction pipeline. Section~\ref{sec:validation} evaluates pipeline performance, and Section~\ref{sec:release} describes the public release and data availability.

\section{Data Processing Architecture}\label{sec:pipeline}

\begin{figure*}[!ht]
    \centering
    \includegraphics[width=0.99\linewidth]{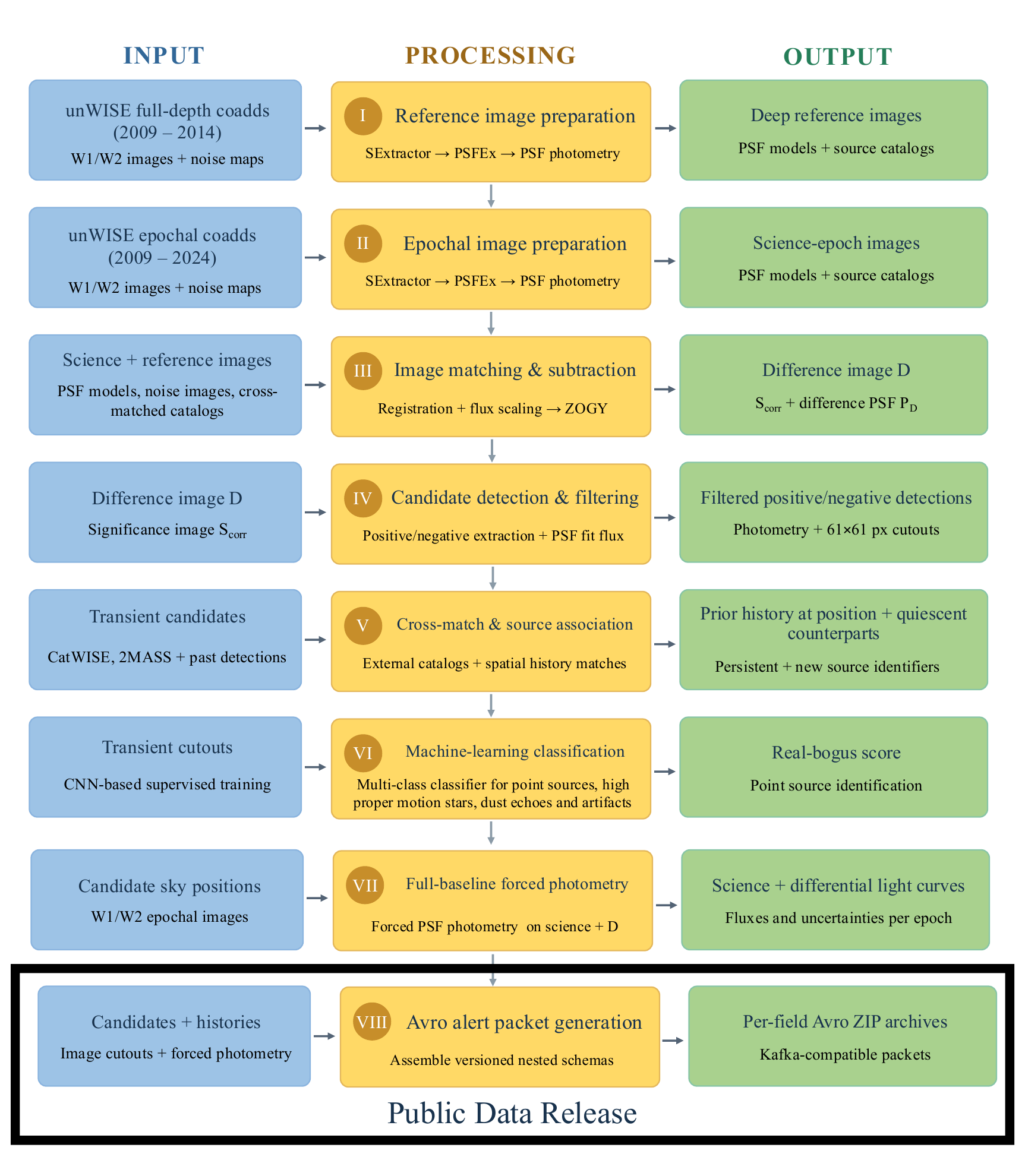}
    \caption{Processing architecture of WTP. We outline eight distinct stages for the pipeline, with each stage using an input data element and producing an intermediate data product that is fed into the following stage; the black rectangle shows the final public data product. The final stage combines all the products into Avro packets distributed as per-field archives.}
    \label{fig:arch}
\end{figure*}

In this section, we describe the processing architecture for WTP starting from the input data to the final delivered products. Figure~\ref{fig:arch} summarizes the eight stages of the pipeline, as described in the following sections. 

\subsection{Input data and processing framework}\label{sec:data}
\label{sec:architecture}

We use \unwise{} co-added images \citep{Lang2014} constructed from calibrated WISE/\neowise{} single-exposure images released via the NASA Infrared Science Archive (IRSA). While the official WISE data products \citep{Wright2010, Mainzer2014} provide images that are convolved with the instrument PSF, unWISE images retain the native spatial resolution of the WISE instrument and avoid correlated noise in downstream image subtraction \citep{Lang2014,De2020}. The WISE survey divides the sky into 18240 image tiles (hereafter `fields') with small overlaps ($\approx 4$\,\arcmin), where each image contains $2048\times2048$ pixels at a scale of $2.75$\,\arcsec/pixel, spanning approximately $1.56^\circ$ on a side. For our pipeline, we use the `full-depth' co-added images using data spanning 2010-2014 as the reference image to subtract against each epoch. Although it contaminates variability detection in the earliest NEOWISE data (from 2014), we use these images due to the improved rejection of artifacts and higher depth in stacks produced by combining the cryogenic mission data (2010-2011) with the first year of NEOWISE \citep{Meisner2017}. 

The time-resolved \unwise{} co-adds provided by \citet{Meisner2018} across the full mission duration are then used as the science images\footnote{Because the reference images were created out of images spanning 2010-2014, difference imaging measurements are affected for transients and variables that were active during this period.}. We refer to each visit to a sky location as an \emph{epoch}, as defined by time-partitioning introduced in \citet{Meisner2018}. Away from the ecliptic poles ($|\beta| < 80^\circ$), an epoch typically combines approximately 12-15 exposures acquired over a few days during a six-month sky pass -- typically resulting in about 24 epochs across most of the sky (numbered as \texttt{e0-e23} in the \unwise\ convention) acquired between UTC 2010-01-07 and 2024-08-01. Each co-added image is accompanied by a standard-deviation map and metadata; because each field occupies a spatially unique footprint in the sky, the pipeline processing was organized by field, band ($W1$ or $W2$), and epoch, with each product from each stage of the pipeline tracked in a central \psql{} database. 

\subsection{Reference image preparation and subtraction}\label{sec:reference}

We first prepare the reference and science images using the same sequence of source extraction and PSF estimation to create inputs required for the image subtraction process. We carry out source extraction using the \sextractor{} package \citep{Bertin1996}, and use the generated source cutouts with the \psfex{} package \citep{Bertin2011} to construct $15\times15$ pixel PSF models. The PSF model is normalized to unit flux and supplied to a second \sextractor{} pass to produce a PSF-photometry catalog. The image, noise map, PSF model, and source catalog are retained as reference products (Stage I). The same procedure is applied to every epochal co-added image (Stage II). Since \unwise{} already supplies calibrated images on a common field grid, we do not repeat the astrometric alignment.

We then match each science image to its reference image to estimate the relative registration and flux-scaling uncertainties for image subtraction. We select a sub-sample of bright, non-blended sources from both PSF catalogs using \texttt{FLAGS=0}, \texttt{FLAGS\_MODEL=0}, and $\texttt{SNR\_WIN}>30$, and matched within $0.5$\,\arcsec ($\lesssim 1/5$ of a pixel). The scatter in their coordinate residuals is used as an estimate of the relative registration uncertainty, while the median ratio of their PSF fluxes is used as the photometric scaling. Both measurements are retained as subtraction metadata. Pixels with zero-valued uncertainties in the input standard-deviation maps are replaced using local values within a five-pixel-wide neighborhood to prevent singular weights. Image subtraction (Stage III) then uses a Python implementation of the ZOGY proper-subtraction formalism of \citet{Zackay2016}. The principal outputs are the difference image $D$, the corrected matched-filter significance image $S_{\rm corr}$, and the effective difference-image PSF $P_D$. Difference-image uncertainties are estimated by propagating the science and reference noise maps and measuring robust background fluctuations (16 - 84 percentile values).

\subsection{Candidate detection and filtering}\label{sec:detection}

We extract candidate variable sources by running \sextractor{} on the masked significance image with a value $S_{\rm corr}>7$ (Stage IV); this yields a detection of all sources that have brightened in the science image relative to the reference image. Since the ZOGY subtraction process identifies all types of statistically significant differences between the science and reference images (including instrument artifacts and other astrophysical non-point sources, e.g., \citealt{Vega2026}), we first implement a cheap filtering layer to remove the worst contaminants. Bright and saturated sources can leave substantial artifacts due to non-linearity and optical diffraction spikes. The pipeline masks these affected regions using variable spatial thresholds for sources with $W1<8$\,Vega\,mag and $W2<7$\,Vega\,mag. We then repeat the extraction by negating the difference image to identify negative transients (fading sources) which are stored identically in the database.

For each detection, we measure the optimal PSF-fit difference flux and its uncertainty, shape parameters (e.g., major and minor axes in pixels, elongation) and PSF goodness of fit metrics. As implemented in the ZTF pipeline \citep{Masci2019}, additional diagnostics include the number of negative pixels surrounding a positive residual and the ratio
\begin{equation}
    sumrat = \frac{\sum D}{\sum |D|},
\end{equation}
evaluated over a local cutout in the difference image $D$. An isolated point-source residual results in $sumrat \approx 1$, whereas approximately balanced positive and negative structures due to dipoles around high proper motion stars result in $sumrat \approx 0$. The initial selection only requires a valid PSF fit flux and an elongation below 2 to remove obvious artifacts; the remaining parameters are recorded in the database for further downstream filtering for specific science applications (see Table \ref{tab:alert}). At this stage we also extract $61\times61$\,pixel ($2.8\arcmin \times 2.8\arcmin$) cutouts of the science, reference, and difference images for inclusion in the alert packets and downstream machine-learning classification \citep{Vega2026}.

\subsection{External cross-matches and real-bogus classification}\label{sec:association}

We cross-match candidates that survive the initial filtering to external catalogs, as well as the internal WTP database to compute history of prior detections (Stage V). Candidates are cross-matched against CatWISE2020 \citep{Marocco2021} and the 2MASS Point Source Catalog \citep{Skrutsie2006} to record up to three neighboring sources within $60\arcsec$, together with the nearest bright CatWISE source within $180\arcsec$\footnote{The astrometric system in WTP is tied to the native \unwise\ system, that is derived from the default WISE astrometric calibration against the 2MASS catalog.}. These associations provide context for identifying counterparts and assessing contamination from nearby bright stars. Candidates are also matched within $5\arcsec$\footnote{We choose this conservatively large cross-match radius despite the better validated astrometric precision due to the observed astrometric scatter near regions of bright nebulosity in the Galactic plane (Section \ref{sec:validation}).} across all epochs prior to the current image to construct the history of prior detections by recording the date of first detection of any candidate at the position and the number of prior detections. These parameters are used for downstream filtering of persistent and eruptive transient sources for scientific applications. 

At this stage, we also evaluate and store a real-bogus score (\texttt{drb}) assessing the likelihood of a candidate to be a point-source using the associated triplet image cutouts (Stage VI). This process uses the convolutional neural network based architecture presented in \citet{Vega2026} that achieves an accuracy of $\approx 93$\% for classification of point sources against instrumental artifacts, extended astrophysical sources (e.g., dust echoes) and dipoles resulting from very high proper motion stars. Visual examination of the resulting scores suggests that the model is very effective at identifying high proper motion stars (which are the largest contaminant outside the Galactic plane), while the separation is poorer for other classes like echoes and artifacts. While varying thresholds on this score produce different levels of false-positive rates, we suggest the use of \texttt{drb}$> 0.1$ to achieve completeness $>99$\% (see \citealt{Vega2026}). 

\subsection{Source association and forced photometry history}\label{sec:fp}

\begin{figure*}[!ht]
    \centering
    \includegraphics[width=0.49\linewidth]{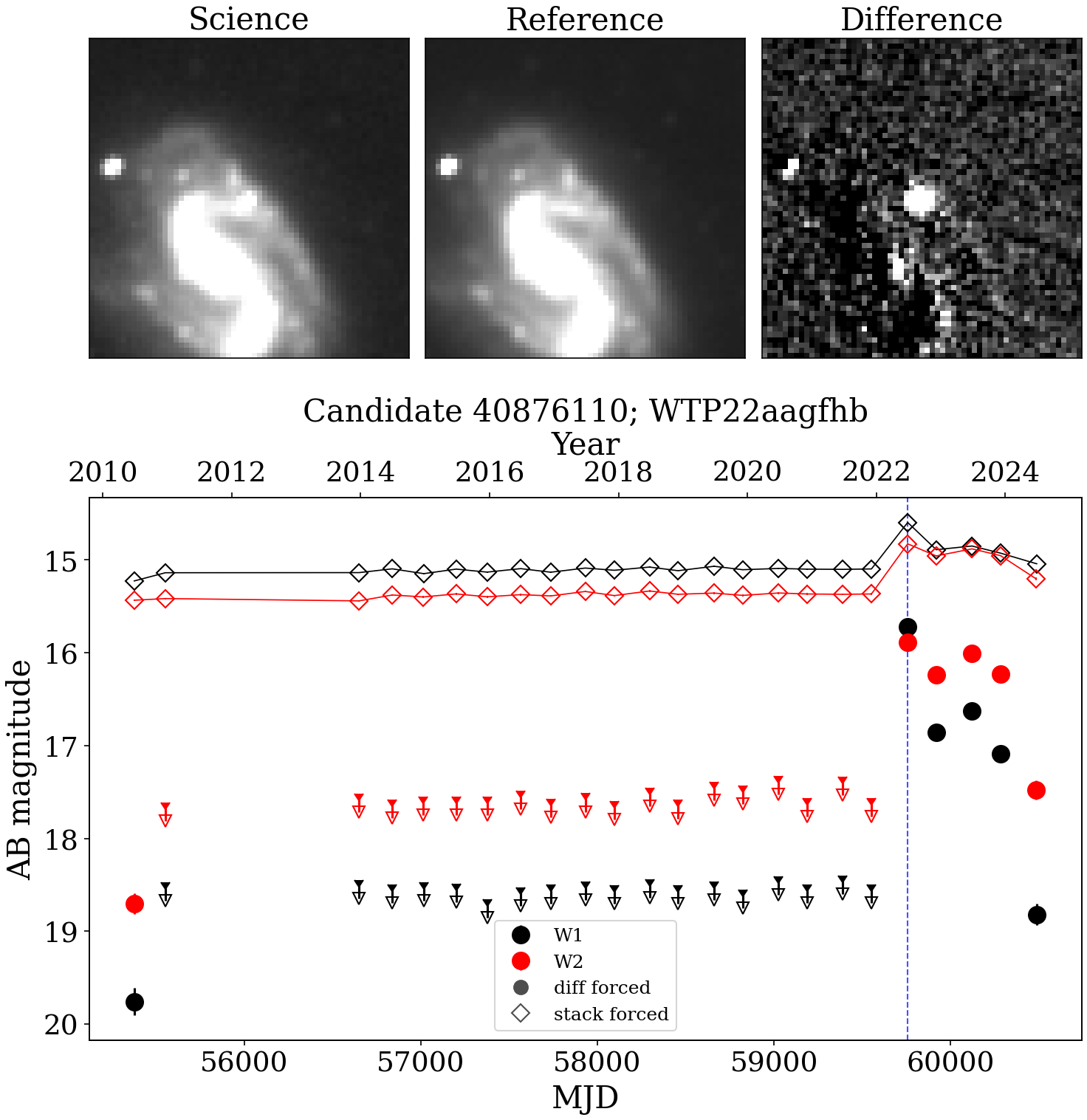}
    \includegraphics[width=0.49\linewidth]{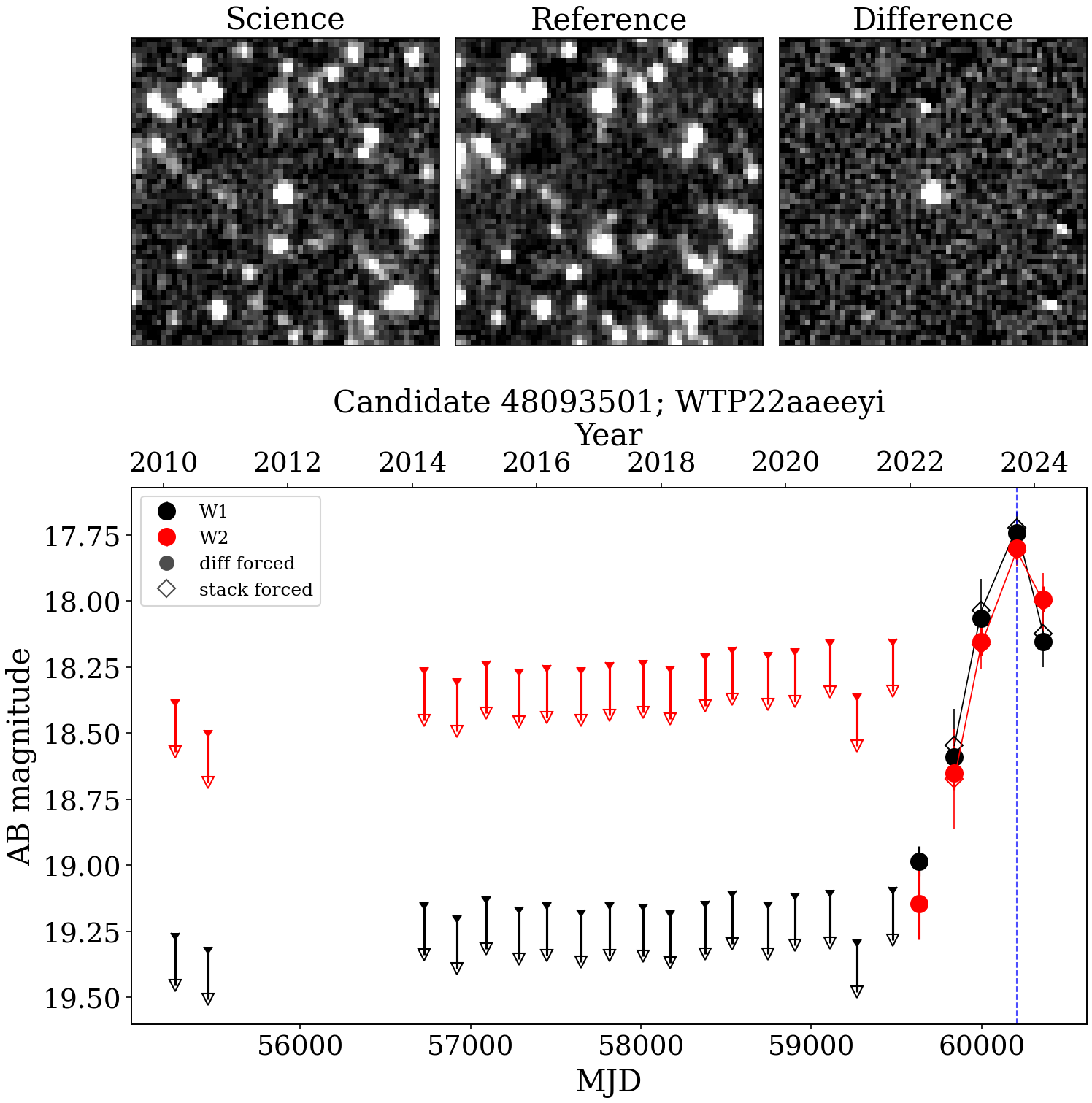}
    \includegraphics[width=0.49\linewidth]{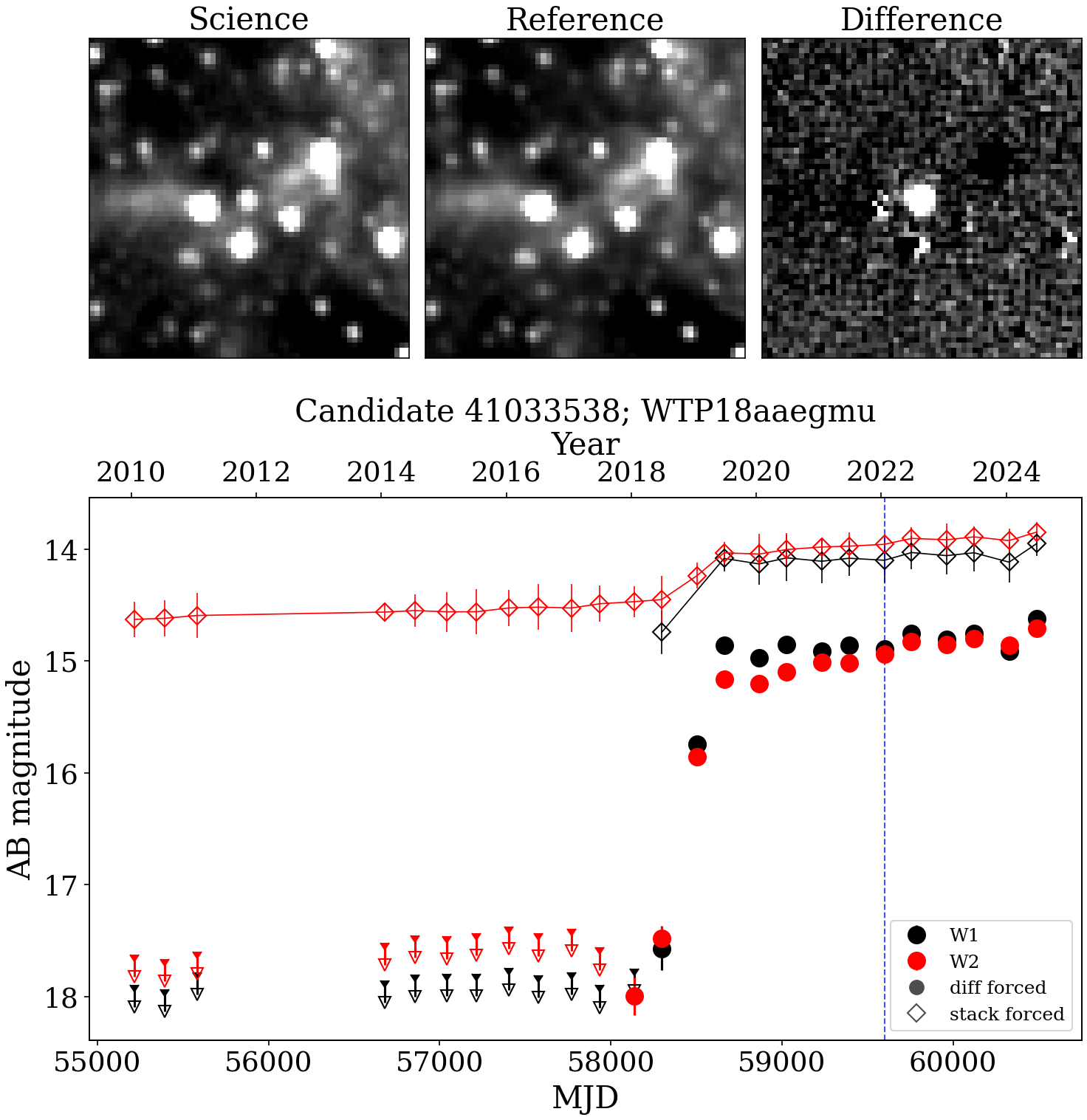}
    \includegraphics[width=0.49\linewidth]{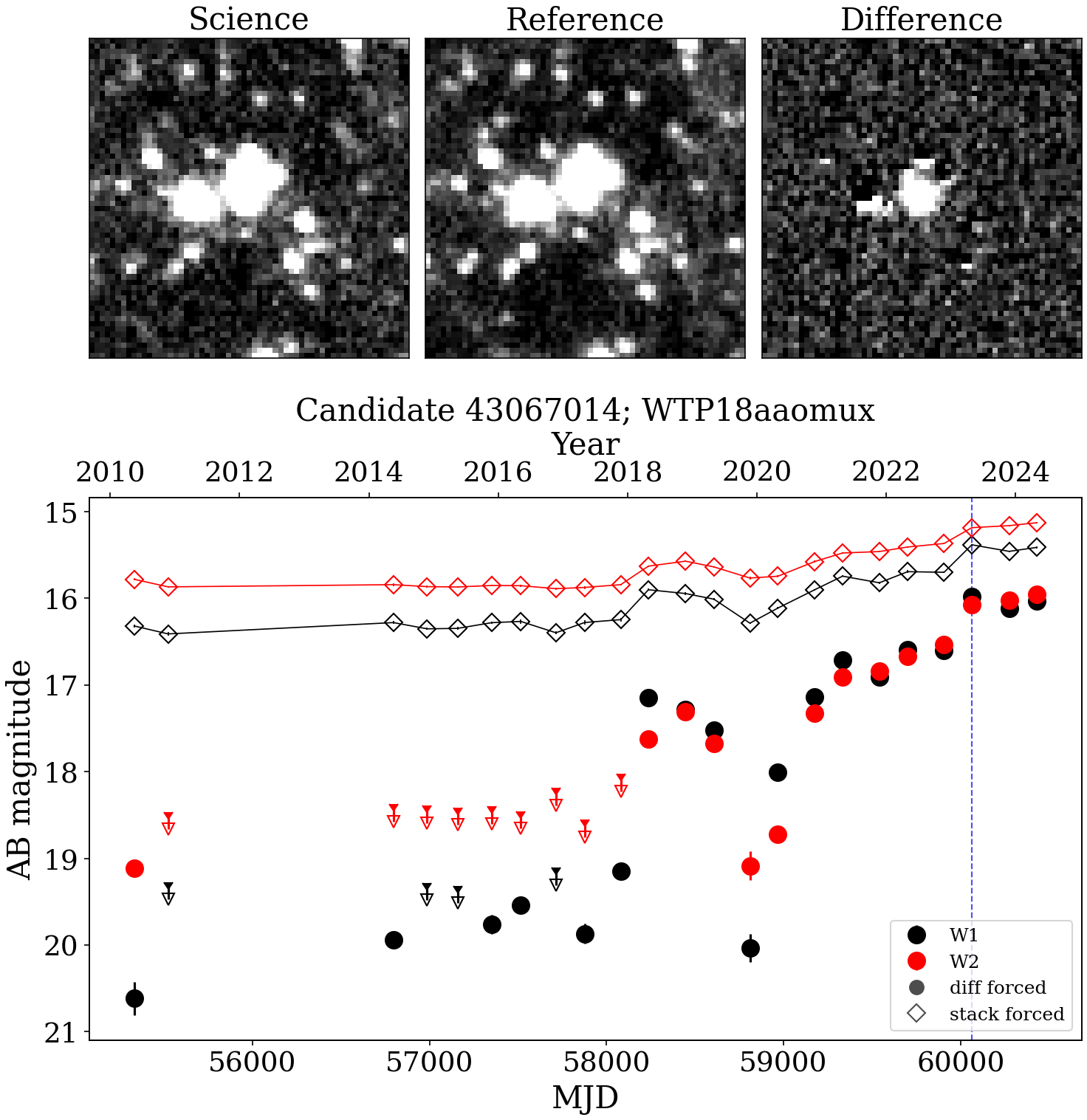}
    \caption{Example light curves of variable and transient sources. In each panel, circles denote forced difference photometry measurements while diamonds denote measurements from the science images. The blue vertical dashed lines show the epoch corresponding to the shown cutouts, and symbols with downward arrows indicate $5\sigma$ upper limits. To demonstrate the diversity of transient environments, we show a typical supernova (WTP\,22aagfhb = SN\,2022jli) on a nearby galaxy (top left), an extragalactic transient with no underlying host (top right), a Galactic plane outburst in a crowded field with nebulosity (bottom left) and a persistent variable Active Galactic Nucleus (bottom right, LEDA 3096435). }
    \label{fig:examples}
\end{figure*}

\begin{figure*}[!ht]
    \centering
    \includegraphics[width=0.99\linewidth]{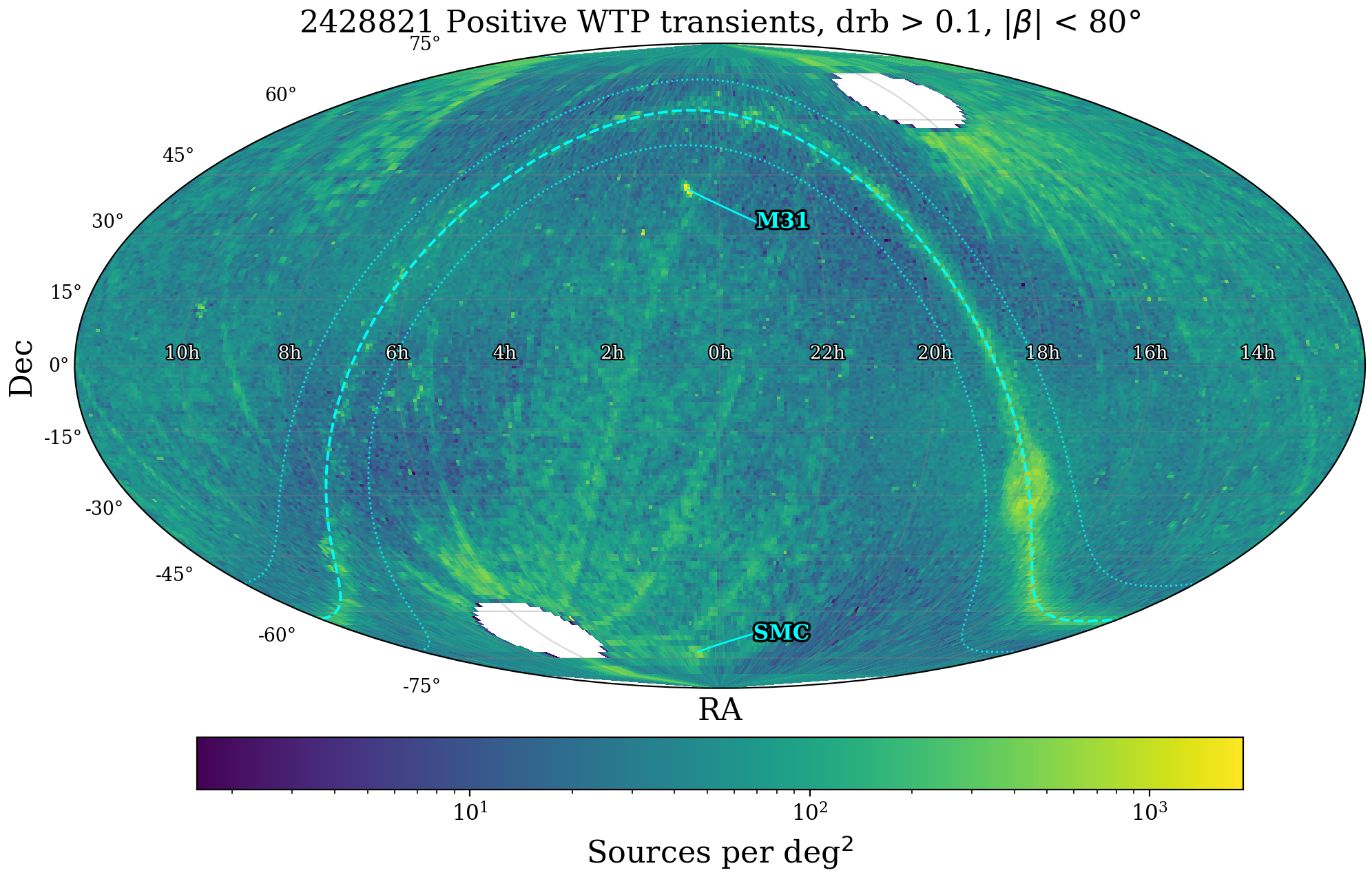}
    \caption{Sky distribution of the source density (color-coded) of WTP variable sources. The Galactic plane is marked with a dashed red line ($b = 0^\circ$) while the central plane ($|b| < 10^\circ$) is highlighted by the red dotted lines. Overdensities associated with nearby galaxies (M31 and SMC) are shown. Regions near the ecliptic pole ($|\beta| > 80^\circ$) are excluded due to the inconsistent image overlaps in this region \citep{Meisner2017}. We observe excess variables tracing the NEOWISE scanning pattern (spikes originating near ecliptic poles), reflecting the inaccuracy of the real-bogus model near field edges \citep{Vega2026}. }
    \label{fig:skydens}
\end{figure*}

The complete list of transient candidates consists of $\approx 50.45$\,M detections across all epochs and parity (positive or negative transients). Many of these transient candidates are spatially associated with the same source, and therefore we assign them under a unique source ID using a spatial cross-match of 5\,\arcsec across the candidate list. Each source is named with a unique WTP ID of the form WTP-YY-XXXXXX where YY is a numeric code denoting the year of first detection (e.g., 14 for 2014) and XXXXXX is an alphabetical incremental code denoting the order of sources detected within a given year (e.g., the first source in 2015 would be WTP15aaaaaa, followed by WTP15aaaaab). The positional grouping results in a catalog of $\approx 18.33$\,M spatially unique variable sources. Due to the abundance of moving variable sources produced by time-variable echoes and high proper motion stars \citep{Vega2026}, the number of static position variable point sources is smaller.  

At this stage, for each candidate source with its corresponding source ID, we construct forced PSF photometry light curves over the complete WISE mission (in both $W1$ and $W2$ bands; Stage VII). We perform the photometry on both the difference images for a given epoch and the original science images. The science image fluxes are particularly useful for characterizing the amplitude of variability relative to the quiescent state, while the difference fluxes fully isolate the variable component. These measurements also enable the identification of low-significance flux detected prior to and after a transient event once a single epoch detection crosses our selection thresholds. For each forced measurement, we compute a formal flux uncertainty from the image noise and an empirical estimate from PSF measurements at 100 random nearby positions. We use the central 16th-84th percentile range of these measurements as an estimate of the photometric uncertainty from the combination of background and nearby confusion noise.

Although not utilizing the alert catalog discussed here, the same forced photometry methods have been used in prior works for targeted searches for variable mid-IR emission \citep{Leung2026, Teja2026, vanRoestel2026, Hillenbrand2026, Fremling2026, Karambelkar2026b, Maeda2026, Rose2026, Rose2025, Mo2025, John2024, Guidry2024, deWit2024, Sharma2023, Hiramatsu2024, Tinyanont2023, Somalwar2023}. In particular, \citet{Leung2026} further improve on the systematics of WISE PSF photometry derived from this pipeline (due to changes in the PSF orientation) by calibrating against non-variable Gaia stars in the same field, allowing high precision time series photometry of high redshift quasars. The total number of forced photometry measurements contained in the packets is $\approx 4.6$\,B ($\approx 50$\,M sources with 2 bands, 23 epochs on average and both science/difference fluxes).

\subsection{Alert generation and final products}\label{sec:alerts}

We assemble the outputs of stages I - VII into flexible Apache Avro\footnote{\url{https://avro.apache.org/}} alert packets for each candidate in the database. Our schema closely follows that implemented for the ZTF survey \citep{Patterson2019} and is listed in Table \ref{tab:alert}. The packet structure is defined by four nested schemas describing the top-level alert (\texttt{Alert}), metadata for the current candidate (\texttt{Candidate}), history of previous candidates at the position (\texttt{Prv\_candidate}), and forced-photometry history over the survey (\texttt{Fp\_record}). Each packet also carries the FITS image cutouts of the science, reference and difference image as compressed gzipped byte streams, together with image metadata, candidate quality diagnostics, and external-catalog associations. Individual Avro packets are bundled into ZIP archives grouped by field for large-scale distribution. The serialization is also compatible with Kafka\footnote{\url{https://kafka.apache.org/}}-based streaming infrastructure, although the archival nature of the analysis only requires one-time ingestion. The complete archive of alerts will be made publicly available as tarballs as well as through the NASA IRSA portal (see Section \ref{sec:release}). Example light curves and cutouts of persistent and eruptive variable sources derived from these packets are shown in Figure \ref{fig:examples}. The sky distribution of all likely real sources (\texttt{drb}$>0.1$) are shown in Figure \ref{fig:skydens}, highlighting the over-densities of variable sources in the Galactic plane and Local Group galaxies.

\section{Pipeline Performance}\label{sec:validation}

\begin{figure*}[!ht]
    \centering
    \includegraphics[width=0.49\linewidth]{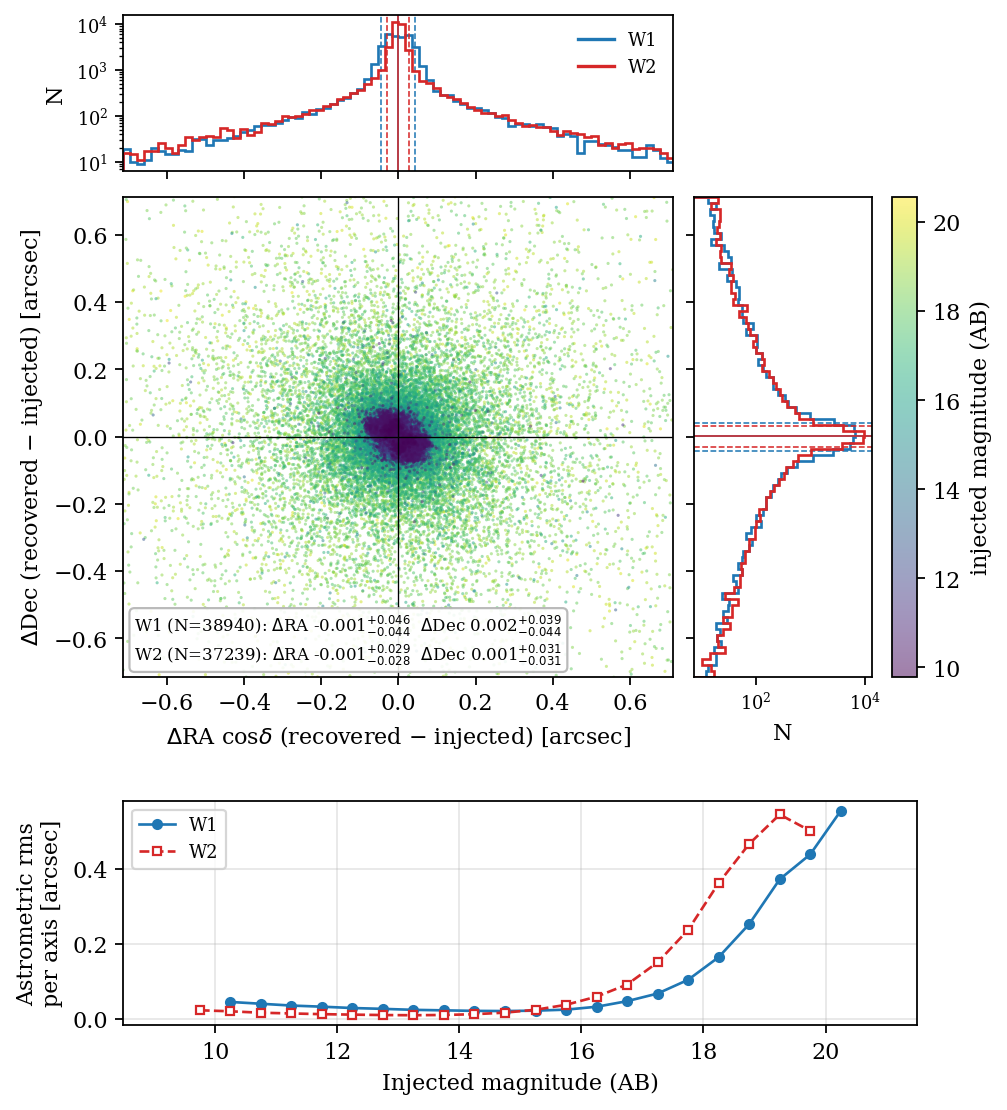}
    \includegraphics[width=0.49\linewidth]{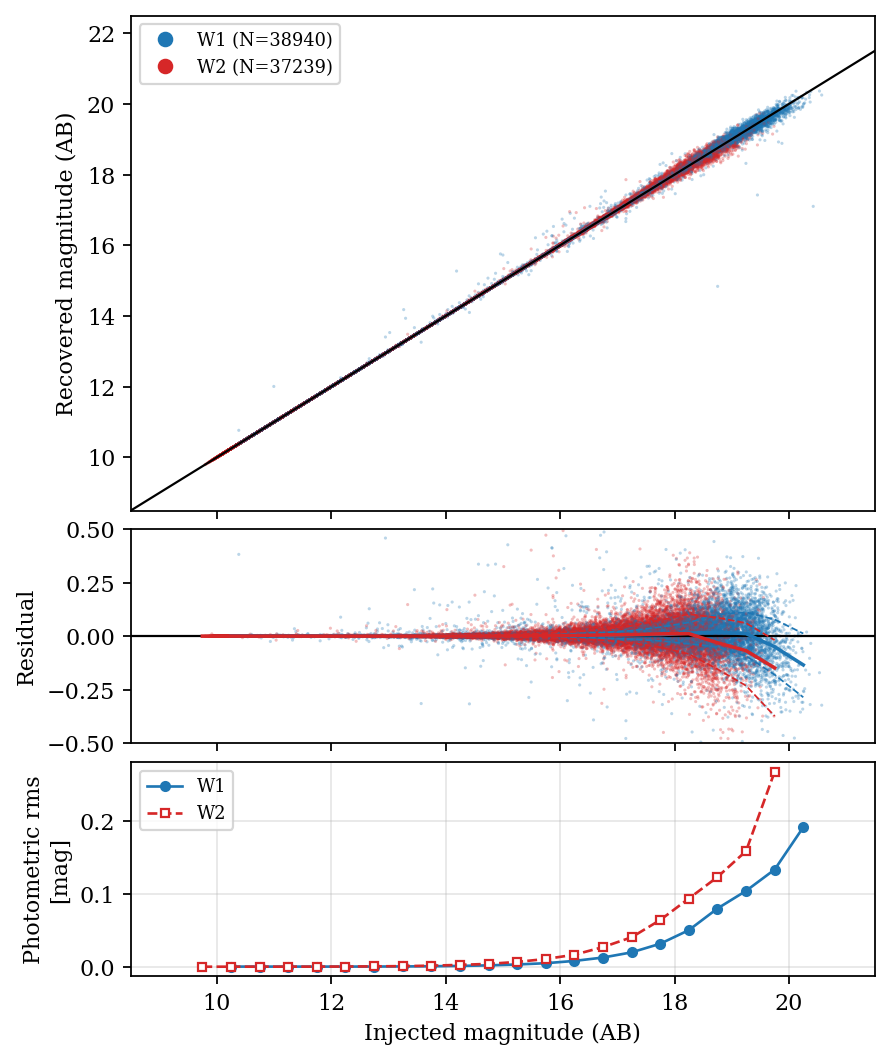}
    \caption{(Left) Astrometric recovery of artificial sources (number indicated in legend). The upper panel shows differences between recovered and injected coordinates, colored by the injected AB magnitude. The marginal logarithmic histograms show the residual distributions, and the inset lists their median offsets and 16th-84th percentiles. The lower panel shows the astrometric rms per coordinate as a function of injected magnitude. (Right) Photometric recovery of artificial sources in $W1$ and $W2$. The upper panel compares recovered PSF magnitudes with injected magnitudes; the black line denotes exact agreement. The middle panel shows the photometry residuals together with their binned trends (solid lines show medians; dashed lines show 16th-84th percentiles). The lower panel shows the photometric standard deviation as a function of injected magnitude.}
    \label{fig:recovery}
\end{figure*}

We characterize the performance of WTP using artificial-source injection and recovery simulations on the \unwise{} coadds -- to measure both astrometric and photometric accuracy of transient detections. Because source recovery varies in crowded Galactic plane fields and in high surface brightness areas given the infrared bandpass and coarse pixel scale \citep{De2020}, we carry out these simulations over a range of Galactic latitudes and extended surface brightness regions of nearby galaxies. For each science image, we construct artificial point sources using the measured science-image PSF, normalized to the desired flux and shifted to the injection position with sub-pixel interpolation. Positions are selected away from image boundaries and regions without valid coverage in either the science or reference image, with a minimum separation between injections to avoid blending the artificial sources with one another.

We carry out fake source injections over the AB magnitude range $\approx 10.0 - 20.0$ ($\approx 7.3 - 17.3$\,Vega\,mag in $W1$ and $\approx 6.7 - 16.7$\,Vega\,mag in $W2$) given the dynamic range of the WISE instrument \citep{Mainzer2014}. We add the sources to the science image while retaining the original reference image, and repeat the image subtraction, bright-source masking, source extraction, and initial candidate filtering described in Section~\ref{sec:pipeline}. A source is considered recovered if a surviving positive detected candidate lies within $5\arcsec$ of its injected position. Figure~\ref{fig:recovery} compares the recovered and injected positions, demonstrating no systematic shifts introduced in the detection process. While the positional accuracy is high for brighter sources ($\lesssim 0.05$\,\arcsec~at $\lesssim 16$\,AB\,mag), the scatter increases to $\approx 0.4-0.5$\,\arcsec~towards fainter sources ($\lesssim 19.5$\,AB\,mag) due to centroiding errors, with the higher scatter for $W2$ at a given magnitude reflecting its overall lower sensitivity. We caution that the astrometric accuracy is degraded relative to these simulations in regions of bright nebulosity (e.g. near Galactic star forming regions) where the high and often variable surface brightness (e.g. due to nearby echoes; \cite{Vega2026}) produce astrometric scatter of up to $1-2$\,pixels. 

\begin{figure*}
    \centering
    \includegraphics[width=0.49\linewidth]{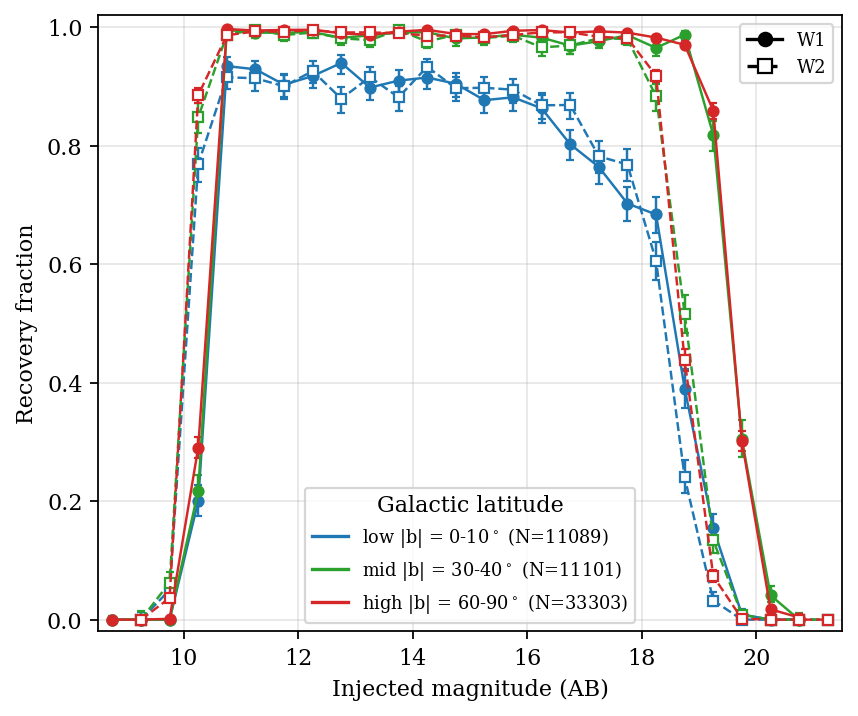}
    \includegraphics[width=0.49\linewidth]{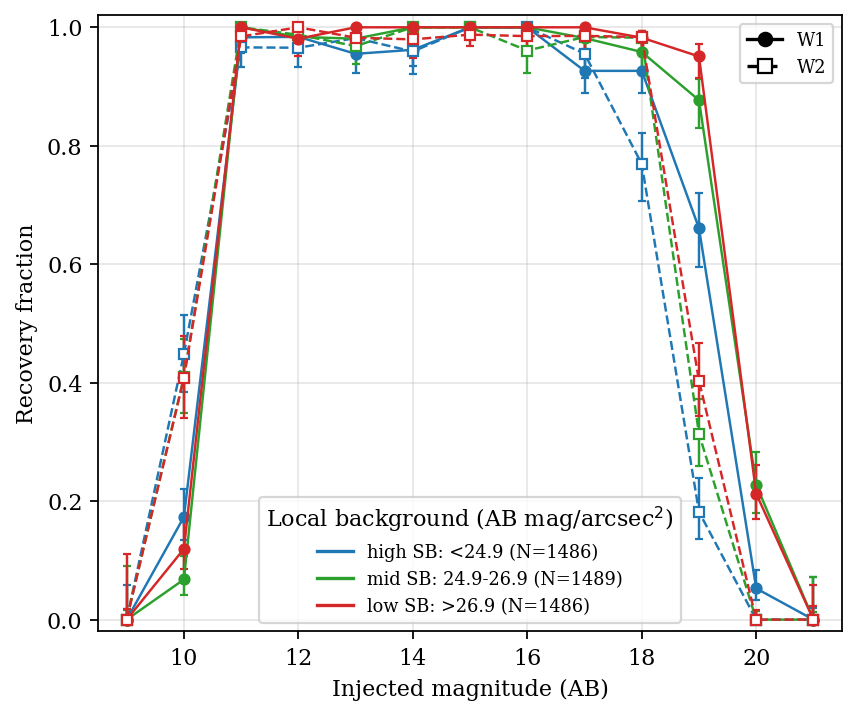}
    \caption{(Left) Detection efficiency as a function of injected AB magnitude for fields in three Galactic latitude intervals (see legend). Error bars show binomial uncertainties on the recovery fractions. (Right) Detection efficiency for artificial sources over a range of local background surface brightness (SB; see legend and text). Error bars are computed as in the left panel.}
    \label{fig:efficiency}
\end{figure*}

Figure~\ref{fig:recovery} also shows a comparison of the recovered PSF magnitudes against the injected values. The recovered fluxes track the input fluxes over the injected magnitude range, with photometric rms below $0.05$\,mag for sources brighter than $\approx 16$\,AB\,mag. The scatter increases toward fainter magnitudes, reaching approximately $20$\% at the sensitivity threshold as expected from Poisson noise. The photometric scatter trend independently confirms a $5\sigma$ limiting magnitude of $\approx 20.1$\,AB\,mag in the $W1$ band and $\approx 19.4$\,AB\,mag in the $W2$ band. The residuals, defined as $\Delta m = m_{\rm recovered}-m_{\rm injected}$, also become systematically negative at the faint end. This behavior is consistent with preferential recovery of sources with nearby bright un-subtracted flux regions that contaminate the photometric measurement. Although difficult to quantify, our tests indicate that the difference photometry can be affected by the rotation of the (somewhat triangular) WISE PSF in very complex backgrounds (e.g. on bright host galaxies) -- producing a systematic alternating (every six months, depending on the WISE scan direction) flux fluctuation that can be corrected with post-processing methods (Lam et al. in prep.).

\subsection{Detection efficiency}

We measure the detection efficiency as the fraction of injected sources recovered in each magnitude bin. Figure~\ref{fig:efficiency} compares this fraction for fields at low ($|b|=0$--$10^\circ$), intermediate ($|b|=30$--$40^\circ$), and high ($|b|=60$--$90^\circ$) Galactic latitudes. Outside the Galactic plane, the recovery fraction is close to unity over a broad magnitude range, before declining rapidly near the sensitivity limit. The intermediate- and high-latitude samples give similar results, with 50\% completeness at approximately 19.6\,AB\,mag in $W1$ and 18.8\,AB\,mag in $W2$, and corresponding 90\% completeness being $\approx 0.5$\,mag shallower in the respective band. At the bright end, recovery drops rapidly for injected magnitudes brighter than $\approx 10.5$\,AB\,mag due to source masking inadvertently rejecting the injected sources. Recovery is lower near the Galactic plane ($|b|<10^\circ$) with the recovery fraction $\approx 90\%$ across much of the magnitude range. The recovery begins to decline at shallower magnitudes than in the higher-latitude samples due to increased confusion in dense stellar regions. The corresponding 50\% completeness limits are approximately 18.6\,AB\,mag in $W1$ and 18.4\,AB\,mag in $W2$; due to the overall lower recovery efficiency, the 90\% completeness limits are much brighter at $\approx 15$\,AB\,mag.

To examine the effect of the local surface brightness on extragalactic transient searches, Figure~\ref{fig:efficiency} also shows the recovery fraction as a function of local background surface brightness. These sources were specifically injected on top of known nearby galaxies ($< 20$\,Mpc) to test source recovery. We estimate the local background from the sigma-clipped median in a $21\times21$ pixel region around each injection position (excluding pixels within a central radius of four pixels), and convert the resulting intensity to surface brightness using the image pixel scale. We categorize the effects of surface brightness by grouping them into three bins of surface brightness. Figure \ref{fig:efficiency} shows that recovery remains high across all three intervals for sufficiently bright sources; however, the recovery fraction decreases more rapidly in regions of higher surface brightness, making the effective recovery limit shallower by up to $\approx 1$\,mag. Although this makes it more challenging to discover faint transients in nearby galaxies, these simulations can be used to correct for this incompleteness in demographic studies (e.g. \citealt{Frohmaier2017, Das2025}). Overall, these simulation results can be used to perform completeness corrections on observed samples of selected transients, i.e. by correcting the observed flux distribution of detected sources for the recovery fractions shown in Figure~\ref{fig:efficiency}.

\section{Public Data Release and Summary}\label{sec:release}

In this paper, we have presented the processing architecture and performance of WTP, a systematic search for transients and variables across nearly 15 years of WISE/\neowise{} observations. Starting from the \unwise{} coadds, we apply proper image subtraction to identify variable (both brightening and fading) sources relative to reference images constructed from the 2010-2014 observations. Forced photometry on the science and difference images provides the full history of each candidate, including epochs when the variable component falls below the nominal pipeline detection threshold. Artificial-source injection and recovery simulations demonstrate positional accuracy of $\lesssim 0.05$\,\arcsec\ and photometric scatter below $0.05$\,mag for sources brighter than $\approx 16$\,AB\,mag, with positional uncertainties increasing to $\approx 0.5$\,\arcsec\ near the faint limit. Outside the Galactic plane, the 50\% completeness limits are $W1 \approx 19.6$ and $W2 \approx 18.8$\,AB\,mag; recovery is reduced in crowded Galactic fields and regions of high surface brightness in nearby galaxies. Our simulations provide a quantitative basis for correcting these incompleteness effects in population studies.

This paper serves as the foundation for the official data release for the WTP catalog -- containing $\approx 50.45$\,M candidate detections of variables associated with $\approx 18.33$\,M spatially unique sources. Each Avro alert packet includes science, reference, and difference image cutouts, candidate photometry and quality diagnostics, a real-bogus classification score, associations with CatWISE and 2MASS sources, and forced-photometry light curves in both $W1$ and $W2$ over the full mission. The complete archive is made available online as a collection of tar-balls organized by the field grid, amounting to a total data volume of $\approx 2.6$\,TB. The catalog is also made available for programmatic queries through the NASA/IPAC Infrared Science Archive (IRSA). 

The discoveries already enabled by WTP demonstrate the scientific reach of an untargeted search for mid-IR variability. This public release makes these searches accessible to the wider community. The diverse range of WTP discoveries has fed into an array of follow-up programs with premier NASA observatories including the Hubble Space Telescope \citep{Uno2026} and James Webb Space Telescope \citep{Masterson2025, De2026b}, setting up the next decade of IR observations with the PRIMA far-IR mission \citep{De2025, Rose2025b}. Although the WISE/\neowise{} observations have ended, their combination of all-sky coverage and mid-IR sensitivity remains a unique resource for the community. In particular, the exquisitely long temporal baseline ($\approx 15$\,yrs) will serve as a legacy historical archive for searches of long-duration events in the era of the Rubin Observatory and Roman Space Telescope. As demonstrated in \citet{Karambelkar2026}, the archival WISE catalogs of slow IR transients are already spearheading discoveries with the SPHEREx mission \citep{Bock2026}. The lessons learnt from the systematic WTP transient searches motivate real-time searches in similar upcoming missions such as the NEO-Surveyor mission \citep{Mainzer2023} which will carry out higher cadence mid-IR surveys of a smaller footprint of the sky.

\section*{Acknowledgements}
We thank S. R. Kulkarni and M. M. Kasliwal for providing feedback throughout the technical and scientific implementation of the project. We thank D. Lang for valuable inputs during the development of the pipeline. We thank Y. Sharma, S. Tinyanont, L. Zuckerman, V. Tran, J. Guidry, J. J. Hermes, C. Myers, C. John, P. Ramesh, P. Nair, C. Fremling, J. van Roestel and D. Stern for valuable feedback at various stages of this project. This publication makes use of data products from the Wide-field Infrared Survey Explorer, which is a joint project of the University of California, Los Angeles, and the Jet Propulsion Laboratory/California Institute of Technology, funded by the National Aeronautics and Space Administration. We acknowledge the support of the National Aeronautics and Space Administration for WISE data analysis through ADAP grant number 80NSSC24K0663 and Roman WFS grant number 80NSSC26K0113. K. D. was supported in part by NASA through the NASA Hubble Fellowship grant \#HST-HF2-51477.001 awarded by the Space Telescope Science Institute, which is operated by the Association of Universities for Research in Astronomy, Inc., for NASA, under contract NAS5-26555. The computations reported in this paper were (in part) performed using resources made available by the Flatiron Institute. During the preparation of this manuscript, the authors used Claude, Opus 5.5, Anthropic to support write code to create figures in this manuscript. The authors reviewed, validated and critically assessed all AI-assisted outputs and take full responsibility for the content of the manuscript.

\startlongtable
\begin{deluxetable*}{lll}
\tablewidth{0pt}
\tabletypesize{\scriptsize}
\tablecaption{Column Names and Descriptions of the WTP Avro Alert Packet (See Text)\label{tab:alert}}
\tablehead{\colhead{Column Name} & \colhead{Data Type} & \colhead{Description}}
\startdata
\multicolumn{3}{l}{\textbf{Alert}} \\
schemavsn & \texttt{string} & Schema version used \\
publisher & \texttt{string} & Origin of alert packet \\
cutoutScience & \texttt{bytes} & Science image cutout (gzipped FITS bytestream) \\
cutoutTemplate & \texttt{bytes} & Reference image cutout (gzipped FITS bytestream) \\
cutoutDifference & \texttt{bytes} & Difference image cutout (gzipped FITS bytestream) \\
objectId & \texttt{string} & WTP source name \\
candid & \texttt{int64} & Candidate ID from operations DB \\
candidate & \texttt{record} & Candidate information \\
prv\_candidates & \texttt{array}\tablenotemark{a} & Prior detections and non-detection upper limits within 5\arcsec, both bands, over the full survey \\
fp\_records & \texttt{array}\tablenotemark{a} & Forced photometry at the candidate position on all difference and science images \\
\hline
\multicolumn{3}{l}{\textbf{Candidate}} \\
mjd & \texttt{float64} & Mid-point MJD of the coadd epoch [days] \\
stackid & \texttt{int64} & Processing ID for science image \\
subid & \texttt{int64} & Processing ID for difference image \\
diffmaglim & \texttt{float32}\tablenotemark{a} & Expected $5\sigma$ mag limit in difference image from global noise estimate [mag] \\
pdiffimfilename & \texttt{string}\tablenotemark{a} & Filename of positive (sci minus ref) difference image \\
candid & \texttt{int64} & Candidate ID from operations DB \\
isdiffpos & \texttt{string} & 1: positive (sci minus ref) subtraction; 0: negative (ref minus sci) \\
epochid & \texttt{int32}\tablenotemark{a} & Epoch ID \\
bandid & \texttt{int32}\tablenotemark{a} & Band ID (1 for $W1$, 2 for $W2$) \\
field & \texttt{int32}\tablenotemark{a} & WTP field ID \\
xpos & \texttt{float32}\tablenotemark{a} & $x$-image position of candidate [pixels] \\
ypos & \texttt{float32}\tablenotemark{a} & $y$-image position of candidate [pixels] \\
ra & \texttt{float64} & Right ascension of candidate, J2000 [deg] \\
dec & \texttt{float64} & Declination of candidate, J2000 [deg] \\
fluxpsf & \texttt{float32} & Flux from PSF-fit photometry [counts] \\
sigmafluxpsf & \texttt{float32} & $1\sigma$ uncertainty in fluxpsf [counts] \\
magpsf & \texttt{float32} & Magnitude from PSF-fit photometry [mag] \\
sigmapsf & \texttt{float32} & $1\sigma$ uncertainty in magpsf [mag] \\
chipsf & \texttt{float32}\tablenotemark{a} & Reduced $\chi^2$ of PSF fit \\
magap & \texttt{float32}\tablenotemark{a} & Aperture magnitude in 4 pixel diameter aperture [mag] \\
sigmagap & \texttt{float32}\tablenotemark{a} & $1\sigma$ uncertainty in magap [mag] \\
fwhm & \texttt{float32}\tablenotemark{a} & FWHM assuming a Gaussian core, from SExtractor [pixels] \\
aimage & \texttt{float32}\tablenotemark{a} & Windowed profile RMS along major axis, from SExtractor [pixels] \\
bimage & \texttt{float32}\tablenotemark{a} & Windowed profile RMS along minor axis, from SExtractor [pixels] \\
elong & \texttt{float32}\tablenotemark{a} & Ratio aimage / bimage \\
nneg & \texttt{int32}\tablenotemark{a} & Number of negative pixels in a $5\times5$ pixel stamp \\
sumrat & \texttt{float32}\tablenotemark{a} & $\sum(\mathrm{pixels}) / \sum|\mathrm{pixels}|$ in a median-filtered $5\times5$ pixel stamp \\
tmdist1,2,3 & \texttt{float32}\tablenotemark{a} & Distance to 1st, 2nd, 3rd closest 2MASS source within 60\arcsec\ [arcsec]\tablenotemark{b} \\
tmmagj1,2,3 & \texttt{float32}\tablenotemark{a} & 2MASS $J$ magnitude of 1st, 2nd, 3rd closest source [mag]\tablenotemark{c} \\
tmmagh1,2,3 & \texttt{float32}\tablenotemark{a} & 2MASS $H$ magnitude of 1st, 2nd, 3rd closest source [mag]\tablenotemark{c} \\
tmmagk1,2,3 & \texttt{float32}\tablenotemark{a} & 2MASS $K_s$ magnitude of 1st, 2nd, 3rd closest source [mag]\tablenotemark{c} \\
wdist1,2,3 & \texttt{float32}\tablenotemark{a} & Distance to 1st, 2nd, 3rd closest unWISE source within 60\arcsec\ [arcsec]\tablenotemark{b} \\
w1mag1,2,3 & \texttt{float32}\tablenotemark{a} & unWISE $W1$ magnitude of 1st, 2nd, 3rd closest source [mag]\tablenotemark{c} \\
w2mag1,2,3 & \texttt{float32}\tablenotemark{a} & unWISE $W2$ magnitude of 1st, 2nd, 3rd closest source [mag]\tablenotemark{c} \\
ndethist & \texttt{int32} & Number of prior detections within 5\arcsec\ in any field and either band\\
mjdstarthist & \texttt{float64}\tablenotemark{a} & Earliest MJD of epoch corresponding to ndethist [days] \\
scorr & \texttt{float64}\tablenotemark{a} & Peak-pixel S/N in matched-filtered detection image \\
mjdstartref & \texttt{float64} & MJD of earliest exposure in reference image [days] \\
mjdendref & \texttt{float64} & MJD of latest exposure in reference image [days] \\
nsubmatches & \texttt{int32} & Number of photometric calibrators matched between science and reference \\
distnearbrstar & \texttt{float32}\tablenotemark{a} & Distance to closest unWISE source with $W1 < 8$ within 180\arcsec\ [arcsec]\tablenotemark{d} \\
magnearbrstar & \texttt{float32}\tablenotemark{a} & $W1$ magnitude of closest unWISE source with $W1 < 8$ within 180\arcsec\ [mag]\tablenotemark{c} \\
drb & \texttt{float32}\tablenotemark{a} & Deep-learning real/bogus score, 0 (bogus) to 1 (real); $-1$ if scoring failed \\
drbversion & \texttt{string} & Real/bogus model version (\texttt{triplet\_v1}; \texttt{failed} if scoring failed) \\
\hline
\multicolumn{3}{l}{\textbf{Prv\_candidate}\tablenotemark{e}} \\
mjd & \texttt{float64} & Mid-point MJD of the coadd epoch [days] \\
subid & \texttt{int64} & Processing ID for difference image \\
stackid & \texttt{int64} & Processing ID for science image \\
diffmaglim & \texttt{float32}\tablenotemark{a} & Expected $5\sigma$ mag limit in difference image [mag] \\
candid & \texttt{int64}\tablenotemark{a} & Candidate ID from operations DB \\
isdiffpos & \texttt{string}\tablenotemark{a} & 1: positive subtraction; 0: negative subtraction \\
epochid & \texttt{int32}\tablenotemark{a} & Epoch ID \\
bandid & \texttt{int32}\tablenotemark{a} & Band ID (1 for $W1$, 2 for $W2$) \\
field & \texttt{int32}\tablenotemark{a} & WTP field ID \\
xpos & \texttt{float32}\tablenotemark{a} & $x$-image position of candidate [pixels] \\
ypos & \texttt{float32}\tablenotemark{a} & $y$-image position of candidate [pixels] \\
ra & \texttt{float64}\tablenotemark{a} & Right ascension of candidate, J2000 [deg] \\
dec & \texttt{float64}\tablenotemark{a} & Declination of candidate, J2000 [deg] \\
magpsf & \texttt{float32}\tablenotemark{a} & Magnitude from PSF-fit photometry [mag] \\
sigmapsf & \texttt{float32}\tablenotemark{a} & $1\sigma$ uncertainty in magpsf [mag] \\
fwhm & \texttt{float32}\tablenotemark{a} & FWHM assuming a Gaussian core, from SExtractor [pixels] \\
scorr & \texttt{float64}\tablenotemark{a} & Peak-pixel S/N in matched-filtered detection image \\
drb & \texttt{float32}\tablenotemark{a} & Deep-learning real/bogus score \\
drbversion & \texttt{string}\tablenotemark{a} & Real/bogus model version \\
\hline
\multicolumn{3}{l}{\textbf{Fp\_record}\tablenotemark{f}} \\
mjd & \texttt{float64} & Mid-point MJD of the coadd epoch [days] \\
subid & \texttt{int64} & Processing ID for difference image \\
stackid & \texttt{int64} & Processing ID for science image \\
diffmaglim & \texttt{float32}\tablenotemark{a} & Expected $5\sigma$ mag limit in difference image [mag] \\
epochid & \texttt{int32}\tablenotemark{a} & Epoch ID \\
bandid & \texttt{int32}\tablenotemark{a} & Band ID (1 for $W1$, 2 for $W2$) \\
field & \texttt{int32}\tablenotemark{a} & WTP field ID \\
xpos & \texttt{float32}\tablenotemark{a} & $x$-image position [pixels] \\
ypos & \texttt{float32}\tablenotemark{a} & $y$-image position [pixels] \\
ra & \texttt{float64}\tablenotemark{a} & Right ascension, J2000 [deg] \\
dec & \texttt{float64}\tablenotemark{a} & Declination, J2000 [deg] \\
forcediffimflux & \texttt{float32} & Forced PSF-fit flux on difference image [counts] \\
forcediffimfluxunc & \texttt{float32} & $1\sigma$ uncertainty in forcediffimflux from random-aperture PSF fits on the image [counts] \\
forcediffimfluxstaterr & \texttt{float32} & $1\sigma$ uncertainty in forcediffimflux from the pixel uncertainty map [counts] \\
forcediffmagpsf & \texttt{float32} & Forced PSF-fit magnitude on difference image [mag]\tablenotemark{g} \\
forcediffsigmapsf & \texttt{float32} & $1\sigma$ uncertainty in forcediffmagpsf [mag]\tablenotemark{g} \\
forcestackimflux & \texttt{float32} & Forced PSF-fit flux on science image [counts] \\
forcestackimfluxunc & \texttt{float32} & $1\sigma$ uncertainty in forcestackimflux from random-aperture PSF fits on the image [counts] \\
forcestackimfluxstaterr & \texttt{float32} & $1\sigma$ uncertainty in forcestackimflux from the pixel uncertainty map [counts] \\
forcestackmagpsf & \texttt{float32} & Forced PSF-fit magnitude on science image [mag]\tablenotemark{g} \\
forcestacksigmapsf & \texttt{float32} & $1\sigma$ uncertainty in forcestackmagpsf [mag]\tablenotemark{g} \\
\enddata
\tablenotetext{a}{Nullable field.}
\tablenotetext{b}{Defaults to 60.0 if no source within 60\arcsec.}
\tablenotetext{c}{Defaults to 99.0 if no source within the search radius.}
\tablenotetext{d}{Defaults to 180.0 if no source within 180\arcsec.}
\tablenotetext{e}{Detections carry all fields. Non-detection rows (an image at this position with no candidate within 5\arcsec) have candid, isdiffpos, xpos, ypos, ra, dec, magpsf, sigmapsf, fwhm, scorr, drb and drbversion set to null; diffmaglim then gives the global limiting magnitude of that image.}
\tablenotetext{f}{One row per epoch and band with a difference image at this position. Epochs where the PSF footprint falls off the image edge are omitted.}
\tablenotetext{g}{Set to $-99$ when the forced flux is $\le 0$.}
\end{deluxetable*}

\bibliography{sample701}{}
\bibliographystyle{aasjournalv7}

\end{document}